\documentclass[pra,twocolumn,showpacs,amsmath,amssymb,aps,a4paper,10pt,floatfix,superscriptaddress]{revtex4}

 \usepackage{graphicx}
 \usepackage{dcolumn}
 \usepackage{bm}
 \usepackage{amssymb,amsmath}
 \usepackage{natbib}
 \usepackage{verbatim}
 \usepackage{subfigure}
 \usepackage{textcomp}
\usepackage[utf8]{inputenc}
\usepackage[T1]{fontenc}
\usepackage{multirow}
\usepackage{booktabs}
\usepackage{ctable}

 \newcommand{\YSO}{Y$_2$SiO$_5$ }
\newcommand{\PRYSO}{Pr$^{3+}$:Y$_2$SiO$_5$ }
\newcommand{\ERYSO}{Er$^{3+}$:Y$_2$SiO$_5$ }
\newcommand{\PRYAG}{Pr$^{3+}$:YAG }
\newcommand{\YSOb}{Y$_2$SiO$_5$}
\newcommand{\PRYSOb}{Pr$^{3+}$:Y$_2$SiO$_5$}
\newcommand{\ERYSOb}{Er$^{3+}$:Y$_2$SiO$_5$}
\newcommand{\PRYAGb}{Pr$^{3+}$:YAG}
\newcommand{\EuYSO}{Eu$^{3+}$:Y$_2$SiO$_5$ }
\newcommand{\EuCl}{EuCl$_3\cdot$6H$_2$O }
\newcommand{\TmYAG}{Tm$^{3+}$:YAG }
\newcommand{\YAG}{YAG }
\newcommand{\zefoz}{ZEFOZ\ }

 \newcommand{\ket}[1]{\left|#1\right\rangle}
 \newcommand{\bra}[1]{\left\langle#1\right|}

\newcommand{\otoprule }{\midrule [\heavyrulewidth ]}

\begin{document}

  \title{Reducing decoherence in optical and spin transitions in rare-earth-ion doped materials}

 \author{D. L. McAuslan}
 \affiliation{Jack Dodd Centre for Photonics and Ultra-Cold Atoms, Department of Physics, University of Otago, Dunedin, New Zealand.}
  \author{J. G. Bartholomew}
 \author{M. J. Sellars}
 \affiliation{Laser Physics Centre, Research School of Physics and Engineering, Australian National University, Canberra, ACT 0200, Australia.}
  \author{J. J. Longdell}
 \email{jevon.longdell@otago.ac.nz}
 \affiliation{Jack Dodd Centre for Photonics and Ultra-Cold Atoms, Department of Physics, University of Otago, Dunedin, New Zealand.}
  \date{\today}
\begin{abstract}

In many important situations the dominant dephasing mechanism in cryogenic rare-earth-ion doped systems is due to magnetic field fluctuations from spins in the host crystal. Operating at a magnetic field where a transition has a zero first-order-Zeeman (ZEFOZ) shift can greatly reduce this dephasing. Here we identify the location of transitions with zero first-order Zeeman shift for optical transitions in \PRYAG and for spin transitions in \ERYSOb. The long coherence times that ZEFOZ can enable would make \PRYAG a strong candidate for achieving the strong coupling regime of cavity QED, and would be an important step forward in creating long-lived telecommunications wavelength quantum memories in \ERYSOb. 
This work relies mostly on published spin Hamiltonian parameters but 
Raman heterodyne spectroscopy was performed on \PRYAG to measure the parameters for the excited state.

\end{abstract}


\maketitle

\section{Introduction}

In a rare-earth-ion-doped system there is a contribution to the homogeneous linewidth from the interaction between the dopants and the atoms that make up the host crystal \cite{equall95}. Fluctuations in the electronic and nuclear spin of the host atoms cause the dopants to experience small random magnetic fields that can alter the transition frequency via the Zeeman effect. This has the effect of broadening the transition. In systems with nuclear spin greater than one applying a magnetic field to the sample produces anticrossings in the energy of the levels, and around these anticrossings it is possible to find transitions that have a zero first-order Zeeman shift \cite{fraval04, fraval05, longdell06,lovric11}. This means that small fluctuations in magnetic field will no longer change the transition frequency, thus decreasing the broadening associated with host-dopant interactions.

The technique of using a \zefoz transition to negate the ion-spin interaction was first used by Fraval \textit{et al.}  in an attempt to lengthen the hyperfine coherence time in \PRYSOb, as it was deemed too short for practical quantum computing applications. They initially demonstrated an increase in phase memory time from $550~\mu$s to 82~ms \cite{fraval04}. By improving the stability of the setup this time was increased to 860~ms \cite{fraval05}. The correlation time for the remaining frequency perturbations was long and the same work used dynamic decoupling techniques to increase echo decay time coherence beyond 10~s. The \zefoz technique has also been used to measure extended coherence times in Pr$^{3+}$:La$_2$(WO$_4$)$_3$ \cite{lovric11}, and the locations of the \zefoz transitions for \EuYSO and \EuCl have been determined \cite{longdell06}.

In this work we provide an extension to the \zefoz technique by showing that it can be applied to optical transitions and the hyperfine transitions of Kramers' ions. 

Reducing the decoherence of an optical transition has previously been investigated in Pr$^{3+}$:LaF$_3$ by Macfarlane \textit{et al.} \cite{macfarlane80}. By driving the nuclear spins of the fluorine atoms with a radio-frequency (RF) field they showed that it was possible to decouple these spins from the Pr$^{3+}$ ions. By choosing the RF field strength such that the angle of the effective field experienced by the fluorine atoms is the `magic angle', the coherence time was increased by an order of magnitude. The \zefoz technique should be more effective at reducing the decoherence because it will reduce the interaction of the dopants with all of the host atoms as opposed to just one type of atom.

As an example, the \zefoz transitions are found for the optical transition at 609.76~nm in \PRYAGb. Based on the measured values of the population lifetime and coherence time in \PRYAGb, it should be possible to increase the optical coherence time by an order of magnitude. If the coherence time can be increased, it will make \PRYAG an excellent candidate for achieving the strong-coupling regime in cavity QED \cite{mcauslan09}. To find a \zefoz transition in \PRYAGb, knowledge of the entire spin Hamiltonian, for both the ground and excited states, is required. Here the necessary spin Hamiltonian parameters are measured using Raman heterodyne spectroscopy.

To date all applications of the \zefoz technique have focused on non-Kramers' ions; however, it would also be useful to determine whether it is possible to find \zefoz transitions for Kramers' ions. As an example, the location of hyperfine \zefoz transitions have been calculated for \ERYSOb. \ERYSO is an interesting system to study because it has the longest optical coherence time of any rare-earth-ion system \cite{bottger09}. If a hyperfine transition could be found that also had a long coherence time, it would make \ERYSO an ideal system for creating a quantum memory at telecommunication wavelengths. Here the location of \zefoz transitions in \ERYSO are calculated using the spin Hamiltonian parameters measured by Guillot-No\"el \textit{et al.} \cite{guillot-noel06}.

\section{Hyperfine Interaction in Rare-Earth-Ions}

The 4f energy levels of a rare-earth-ion are completely described by the following Hamiltonian \cite{Book-spectroscofsolidsrareearth}:
\begin{equation} \label{eq:spinham1}
H = H_{FI} + H_{CF} + H_{HF} + H_Q + H_{eZ} + H_{nZ},
\end{equation}
where the first two terms represent the free-ion and crystal-field Hamiltonians and determine the electronic energy level structure. The other four terms are much smaller than the first two and describe the hyperfine structure of the energy levels. These terms are the hyperfine, quadrupole, electronic Zeeman and nuclear Zeeman Hamiltonians respectively. As the free-ion energy levels are well known, and only the lowest crystal-field level is populated at cryogenic temperatures, only the final four terms of Eq.~(\ref{eq:spinham1}) need to be considered when calculating the energy of hyperfine levels.

In this work two different materials are considered; \PRYAG which is a non-Kramers' ion (has an even number of f-electrons), and \ERYSO which is a Kramers' ion (odd number of f-electrons). Different Hamiltonians are used to describe the two systems.

For a non-Kramers' ion in a site of low symmetry the electronic angular momentum is quenched, lifting the degeneracy of the electronic states. Applying second-order perturbation theory to the system causes second-order hyperfine and second-order electronic Zeeman terms to arise that are of a similar magnitude to the nuclear Zeeman and quadrupole interactions. Therefore, non-Kramers' ions can be described by the effective Hamiltonian derived by Teplov \cite{teplov68}:
\begin{align} \label{eq:hamteplov}
H_{\text{eff}} &= g_J^2\mu_B^2 \boldsymbol{B}\cdot \boldsymbol{\Lambda} \cdot \boldsymbol{B} + \boldsymbol{B} \cdot (2 A_J  g_J \mu_B \boldsymbol{\Lambda} + \hbar  \gamma_N) \cdot \boldsymbol{I} \nonumber \\
& \quad + \boldsymbol{I} \cdot (A_J^2 \boldsymbol{\Lambda} + \boldsymbol{T}_Q) \cdot \boldsymbol{I},
\end{align}
where $g_J$ is the Land\'{e} g value,  $\mu_{B}$ is the Bohr magneton, $\gamma_N$ is the nuclear gyromagnetic ratio, and $A_J$ is the magnetic hyperfine constant. $\boldsymbol{B}$ is the applied magnetic field, $\boldsymbol{I}$ is a vector of the nuclear angular momentum operators, and $\boldsymbol{\Lambda}$ is a tensor that describes the second-order interactions in the system. 

The term $ g_J^2\mu_B^2 \boldsymbol{B}\cdot \boldsymbol{\Lambda} \cdot \boldsymbol{B}$ describes the quadratic electronic Zeeman interaction \cite{Book-spectroscofsolidsrareearth}. $ \boldsymbol{B} \cdot (2 A_J  g_J \mu_B \boldsymbol{\Lambda} + \hbar  \gamma_N) \cdot \boldsymbol{I}$ is the enhanced nuclear Zeeman interaction, where the nuclear Zeeman interaction is enhanced due to coupling to the second-order hyperfine interaction \cite{bleaney73, abragam83}. $\boldsymbol{I} \cdot (A_J^2 \boldsymbol{\Lambda} + \boldsymbol{T}_Q) \cdot \boldsymbol{I}$ is the combined quadrupole interaction, made up of the second-order hyperfine interaction and the pure-quadrupole interaction. Because the second-order hyperfine interaction can be written in the same form as the pure-quadrupole interaction, it is often called the pseudo-quadrupole interaction \cite{baker58}.

For the work done on \PRYAG the dopants occupy sites of sufficiently high symmetry that the axes of the interaction tensors coincide with the site axes. This enables the Hamiltonian to be diagonalized, i.e.
\begin{equation}\label{eq:hampr}
 H_{\text{eff}} = \boldsymbol{B} \cdot \boldsymbol{Z} \cdot \boldsymbol{B} + \boldsymbol{B} \cdot \boldsymbol{M} \cdot \boldsymbol{I} + \boldsymbol{I} \cdot \boldsymbol{Q} \cdot \boldsymbol{I},
\end{equation}
where $\boldsymbol{M}$ is the enhanced nuclear Zeeman tensor:
\[ \boldsymbol{M} = \begin{bmatrix}
 \gamma_x & 0 & 0 \\
 0 &\gamma_y & 0 \\
 0 & 0 & \gamma_z \end{bmatrix},\] 
$\boldsymbol{Q}$ is the tensor of combined quadrupole constants:
\[ \boldsymbol{Q} = \begin{bmatrix}
 E & 0 & 0 \\
 0 & -E & 0 \\
 0 & 0 & D \end{bmatrix},\]
and $\boldsymbol{Z}$ is the second-order Zeeman tensor:
\[ \boldsymbol{Z} = \begin{bmatrix}
 Z_x & 0 & 0 \\
 0 & Z_y & 0 \\
 0 & 0 & Z_z \end{bmatrix}.\]

When considering ions with an odd number of electrons (such as Er$^{3+}$), Kramers' theorem states that all levels will have an electronic degeneracy that can only be lifted by applying a magnetic field to the sample \cite{Book-spectroscofsolidsrareearth}. This results in a large contribution to the Hamiltonian from the first-order electronic Zeeman and magnetic hyperfine interactions. At cryogenic temperatures only the lowest crystal-field level is populated; therefore, the system can be approximated as having spin $S = \frac{1}{2}$ \cite{guillot-noel06}. The enhanced nuclear Zeeman and combined quadrupole interactions are as for non-Kramers' ions. Thus, the energy levels of Kramers' ions can be described by the following effective Hamiltonian \cite{Book-spectroscofsolidsrareearth}:
\begin{align} \label{eq:hamkramers}
H_{\text{eff}} &= \mu_B \boldsymbol{B} \cdot \boldsymbol{g} \cdot \boldsymbol{S} + \boldsymbol{I} \cdot \boldsymbol{A} \cdot \boldsymbol{S} + \boldsymbol{I} \cdot (A_J^2 \boldsymbol{\Lambda} + \boldsymbol{T}_Q) \cdot \boldsymbol{I} \nonumber \\
& \quad -  \boldsymbol{B} \cdot (2 A_J  g_J \mu_B \boldsymbol{\Lambda} + \hbar  \gamma_N) \cdot \boldsymbol{I},
\end{align}
where $\mu_B \boldsymbol{B} \cdot \boldsymbol{g} \cdot \boldsymbol{S}$ is the electronic Zeeman term, and $\boldsymbol{I} \cdot \boldsymbol{A} \cdot \boldsymbol{S}$ is the first-order hyperfine interaction.

To simplify the description for \ERYSO we will rewrite the above Hamiltonian as \cite{guillot-noel06}:
\begin{equation}\label{eq:erham}
H_{\text{eff}} = \mu_B \textbf{B} \cdot \textbf{g}\cdot \textbf{S} + \textbf{I}\cdot \textbf{A} \cdot \textbf{S} + \textbf{I}\cdot \textbf{Q}\cdot \textbf{I} - \mu_n g_n \textbf{B}\cdot \textbf{I},
\end{equation}
where $\mu_n$ is the nuclear magneton and $g_n$ is the nuclear g-factor = -0.1618 for $^{167}$Er . $\textbf{g}$, $\textbf{A}$, and $\textbf{S}$ are the g-factor matrix, matrix of hyperfine parameters and the electronic spin matrix respectively. 

\section{Pr$^{3+}$:YAG}

Y$_3$Al$_5$O$_{12}$ (YAG) has a cubic crystal structure. The praseodymium ions substitute for yttrium ions that are located at dodecahedral $c$-sites, and have local $D_2$ symmetry \cite{wang97, sun00}. This means there are six different orientations that the praseodymium ions, which substitute for yttrium, can have in the crystal. This is explained in more detail in the context of \TmYAG in \cite{sun00}. The six sites are magnetically inequivalent, meaning the ions' hyperfine levels will experience a site-dependent frequency shift under the application of an external magnetic field. The orientation of the local $x$-, $y$-, $z$-axes of site 1 are defined as being along the [1,1,0], [1,-1,0] and [0,0,1] directions of the crystal. The orientations of the other sites are found by performing an appropriate rotation on the site 1 axes.

The transition of interest is between the $^1$D$_2$ excited state and $^3$H$_4$ ground state in \PRYAGb, occurring at 609.76~nm (in vacuum) \cite{macfarlane96}. The inhomogeneous linewidth of this transition has been measured as 1 -- 2 cm$^{-1}$ \cite{shelby83,macfarlane96,wang97} and has an oscillator strength of $f=1.3 \times 10^{-6}$ \cite{macfarlane_private}. Shelby \textit{et al.} measured an optical coherence time of $T_{2(opt)}=20~\mu$s \cite{shelby83} compared to an excited state lifetime of $T_{1(opt)} = 230~\mu$s \cite{wang97}. They put the discrepancy down to magnetic dephasing caused by spins in the host. The coherence can be expected to be shorter for \PRYAG than for example \PRYSO because the nuclear magnetic moment of aluminium (3.64 \cite{kaye95}), the dominant spin in \YAG is much larger than for yttrium (-0.137 \cite{kaye95}), the dominant spin in \YSOb.

If this excess dephasing is purely magnetic, it should be easily reduced to a level where the coherence time is lifetime limited, requiring only a twenty fold increase in coherence time to reach a $T_2=2T_1$. This can be compared to the thousand fold increases demonstrated for hyperfine transitions. Often in rare earth doped systems the broadening due to phonon scattering is negligible at cryogenic temperatures \cite{Book-spectroscofsolidsrareearth}; however, there is a possibility that this could be a more significant issue with \PRYAG because of a crystal field level not far from the ground state \cite{wang97}.

The hyperfine structure of Pr$^{3+}$:YAG was first studied by Shelby \textit{et al.} where spectral hole burning was used on a 0.15\% Pr$^{3+}$:YAG crystal to measure the hyperfine splittings of the ground and excited states \cite{shelby83}. Pr$^{3+}$ has zero electronic spin and a nuclear spin of 5/2, resulting in three degenerate pairs of hyperfine levels for both the ground and excited states. Wang later confirmed these hyperfine splittings using optically detected nuclear magnetic resonance \cite{wang97}, and they also agree with what we have measured using Raman heterodyne spectroscopy. 

An in-depth investigation into the hyperfine structure of \PRYAG was performed by Wang \cite{wang97}. Based on the spin Hamiltonian of the system and the measured hyperfine splittings, he determined the values of the effective quadrupole tensor for both the ground and excited states. Measurements were also performed using Raman heterodyne spectroscopy on the 33.33~MHz ground-state hyperfine transition. By observing how the transition energies varied with magnetic field, the values of the enhanced nuclear Zeeman tensor were calculated. The known spin Hamiltonian parameters are summarized in Table~\ref{tab:t_pryag2}.

\begin{table}[t]
\caption{\label{tab:t_pryag2} Spin Hamiltonian parameters measured by Wang \cite{wang97}.}
\begin{center}
\begin{tabular}{ccc}
\toprule
\toprule
 & Ground State ($^3$H$_4$) & Excited State ($^1$D$_2$)  \\
\otoprule
 $ \delta_1$ (MHz) & 33.33 & 6.41 \\
 $ \delta_2$ (MHz) & 41.86 & 8.32  \\
 $D$ (MHz) & 11.36 & 2.24 \\
$E$ (MHz) & 2.776 & 0.520 \\
$\gamma_x$ (kHzG$^{-1}$) & 13 & -- \\
$\gamma_y$ (kHzG$^{-1}$) & 1.6 & -- \\
$\gamma_z$ (kHzG$^{-1}$) & 30.3 & -- \\
\bottomrule
\bottomrule
\end{tabular}
\end{center}
\end{table}

To find an optical \zefoz transition in \PRYAGb, knowledge of the entire spin Hamiltonian for both the ground and excited states is required. Therefore, the excited-state enhanced nuclear Zeeman tensor needs to be measured. This can be done using Raman heterodyne spectroscopy \cite{mlynek83, wong83}.

\subsection{Measuring Spin Hamiltonian Parameters}

\begin{figure}[t]
  \begin{center}
  \includegraphics[width=0.48\textwidth]{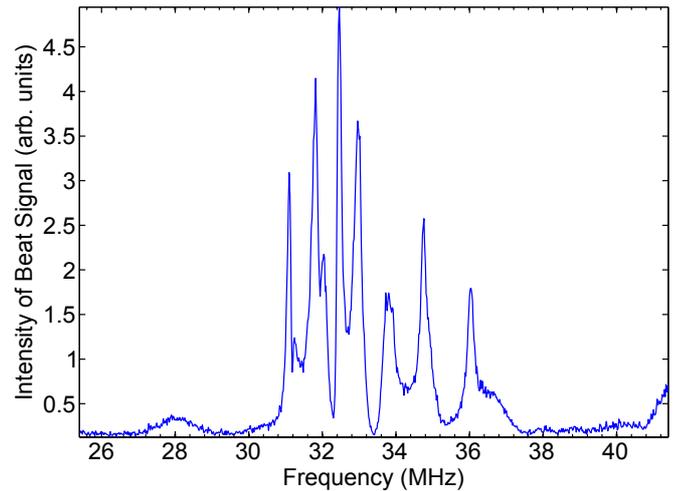}
\end{center}
\caption{\label{fig:ramanspectra} (Color online) Example Raman heterodyne spectrum centred on the 33.33~MHz ground-state transition. A beat is detected when the RF field applied to the sample is resonant with a hyperfine transition.}
\end{figure}

Raman heterodyne spectra were obtained using an experimental setup similar to that described in reference \cite{longdell02}. An upgraded set of Helmholtz coils was used that enabled DC magnetic fields of up to 200~G to be generated. When recording the spectra the sample was immersed in super-fluid liquid helium to produce temperatures less than 2.2~K as this gave the strongest Raman heterodyne signal (most likely due to the sample being colder). Figure \ref{fig:ramanspectra} is an example of the recorded spectra.

The crystal used in the experiments was a $0.5\%$ Pr$^{3+}$:YAG crystal with an initially unknown orientation, grown at the Australian National University. The crystal was placed in the cryostat so that the crystal faces are perpendicular to the axes of the Helmholtz coils. Using Raman heterodyne spectroscopy on the 33.33~MHz ground-state transition (the same transition used by Wang to measure the ground-state parameters), the crystal orientation was determined to be that the $X$-, $Y$-, $Z$-axes are approximately aligned along the [1,-1,0], [1,1,1] and [1,1,-2] directions.

\begin{figure}[t]
 \begin{center}
  \includegraphics[width=.48\textwidth]{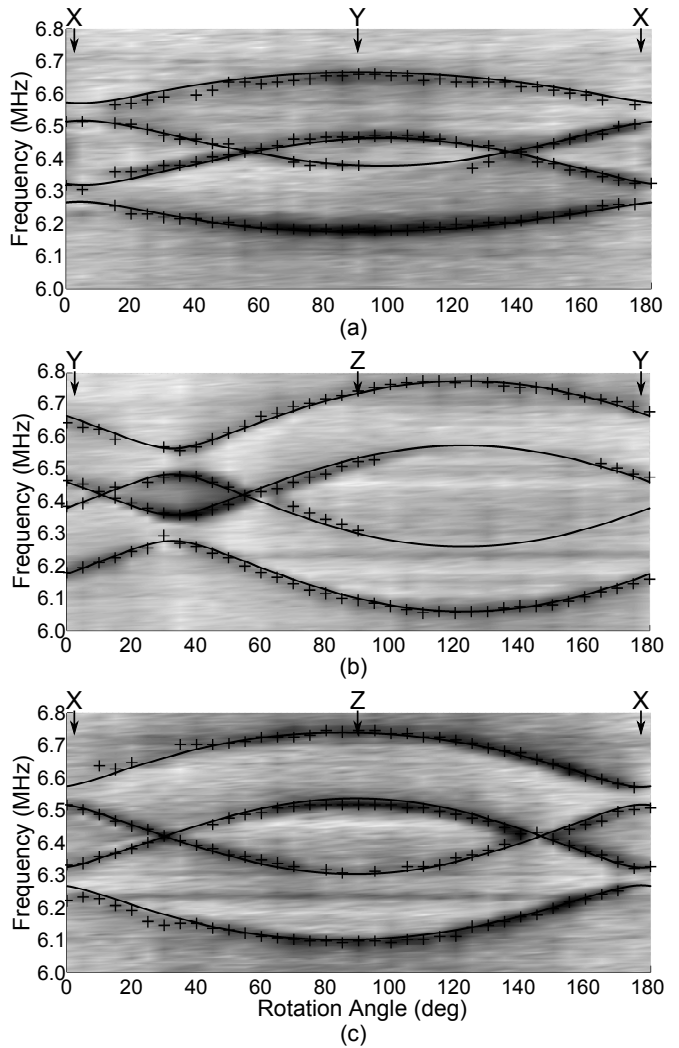}
\end{center}
\caption{\label{fig:excitedstate} Raman heterodyne spectra obtained for the excited-state hyperfine transition in Pr$^{3+}$:YAG centred on 6.41~MHz. The horizontal axis corresponds to the orientation of the applied DC magnetic field (B$_0 = 68 \pm 2$~G) , the vertical axis the frequency of the applied RF, and the shades of gray represents the intensity of the beat signal. Black crosses are the peaks selected from each individual spectrum. Solid lines are the best fit to the crosses. The spectra correspond to the magnetic field being rotated about the; (a) $X$-$Y$ plane, (b) $Y$-$Z$ plane, and (c) $X$-$Z$ plane.}
\end{figure}

Raman heterodyne spectroscopy was then performed on the 6.41~MHz transition so that the components of the excited-state enhanced nuclear Zeeman tensor can be determined. The recorded spectra are shown in Fig.~\ref{fig:excitedstate}. Each plot corresponds to a series of spectra (each one similar to Fig.~\ref{fig:ramanspectra}) laid side by side, each with a different DC magnetic field applied to the sample. The magnetic field coils perform a rotation of 180$^{\circ}$ around the $X$-$Z$ plane, 180$^{\circ}$ around the $X$-$Y$ plane, and 180$^{\circ}$ about the $Y$-$Z$ plane ($X$, $Y$, $Z$ directions are defined in terms of the three sets of Helmholtz coils). Due to the weakness of this transition, only four of the twenty-four possible lines were visible. Each measured spectrum was inspected manually and the frequencies of the peaks corresponding to transitions were selected. These frequencies are represented by the black crosses in Fig.~\ref{fig:excitedstate}.  

To calculate the excited state nuclear Zeeman parameters a computer program was run that minimized the misfit between the measured data and the theoretical values (calculated from the spin Hamiltonian). There were two extra factors that needed to be taken into consideration when performing the fit. Firstly, an extra rotation was applied to the spin Hamiltonian to allow for imperfect alignment between the sample and the coil axes. This extra rotation is defined as:
\[R_s = \begin{bmatrix}
 1 & 0 & 0 \\
 0 & \text{cos} \alpha & \text{sin} \alpha \\
 0 & -\text{sin} \alpha & \text{cos} \alpha \end{bmatrix} \begin{bmatrix}
 \text{cos} \beta & 0 & -\text{sin} \beta \\
 0 & 1 & 0 \\
 \text{sin} \beta & 0 & \text{cos} \beta \end{bmatrix} \begin{bmatrix}
 \text{cos} \gamma & \text{sin} \gamma & 0 \\
 -\text{sin} \gamma & \text{cos} \gamma & 0 \\
 0 & 0 & 1 \end{bmatrix},  \] 
The second consideration was that the principle axes of the combined quadrupole tensor are not necessarily the same for the ground and excited states. However, due to the local $D_2$ symmetry of the praseodymium ions, the assumption was made that the axes of the combined quadrupole tensor should lie along the three $C_2$-axes ([1,1,0], [1,-1,0], and [0, 0, 1] for site 1). The best fit was found for:
\[
\boldsymbol{Q}_{\text{ex}} =\begin{bmatrix}
 D & 0 & 0 \\
 0 & E & 0 \\
 0 & 0 & -E \end{bmatrix},
\]
and is represented by the solid lines in Fig.~\ref{fig:excitedstate}. The parameters of this fit are given in Table~\ref{tab:t_pryag3}.

\begin{table}[t]
\caption{\label{tab:t_pryag3} Parameters of the best fit to the data obtained for the 6.41~MHz excited-state transition of \PRYAGb.}
\begin{center}
\begin{tabular}{cc}
\toprule
\toprule
Excited-State Parameter & Value  \\
\otoprule
$\gamma_x$ (kHzG$^{-1}$) & 1.39  \\
$\gamma_y$ (kHzG$^{-1}$) & 1.68  \\
$\gamma_z$ (kHzG$^{-1}$) & 1.62  \\
$\alpha$ & -2.7$^{\circ}$  \\
$\beta$ & 0.4$^{\circ}$  \\
$\gamma$ & 3.8$^{\circ}$  \\
\bottomrule
\bottomrule
\end{tabular}
\end{center}
\end{table}

When fitting to the Raman heterodyne data the second-order Zeeman term ($\boldsymbol{B} \cdot \boldsymbol{Z} \cdot \boldsymbol{B}$) was excluded from the calculations. While this term adds a frequency shift dependent on the magnitude of the applied magnetic field, the shift is constant for all hyperfine levels of a particular state. Therefore, the frequency of the hyperfine transitions is independent of this term. However, it is necessary to include the second-order Zeeman term when calculating the relative frequencies of optical transitions.

\subsection{Finding the \zefoz transitions}

The motivation behind measuring the spin Hamiltonian parameters for \PRYAG is to find an optical transition where there is zero first-order Zeeman shift, resulting in an extended coherence time. Before calculating what magnetic field to apply to the sample, it is necessary to calculate the components of the second-order Zeeman tensor, $ \boldsymbol{Z} = g_J^2 \mu_B^2 \boldsymbol{\Lambda}$. The second-order hyperfine tensor is calculated from second-order perturbation theory \cite{Book-spectroscofsolidsrareearth}:
\begin{equation} \label{eq:Lambda2}
 \Lambda_{\alpha \beta} = \displaystyle\sum\limits_{n=1}^{2J+1} \frac{\bra{0} J_\alpha \ket{n} \bra{n} J_\beta \ket{0}}{E_n - E_0}
\end{equation}
where $\ket{0}$ is the level of interest, $\ket{n}$ refers to the other crystal field levels, and $E$ is the energy of the level. Here $\boldsymbol{\Lambda}$ is calculated by subtracting the nuclear gyromagnetic ratio ($\gamma_N = 1.304$~kHzG$^{-1}$ for praseodymium) from its enhanced value ($\gamma_{\alpha} = \gamma_x, \gamma_y, \gamma_z$), that is:
\begin{equation} \label{eq:Lambda}
 \Lambda_{\alpha \alpha} =  \frac{\left( \gamma_{\alpha}-\gamma_N \right) \hbar}{2 g_J \mu_B A_J},
\end{equation}
where the magnetic hyperfine constant is given by:
\begin{equation}
A_J = 2 \mu_B \gamma_N \hbar \langle r^{-3} \rangle \langle J || N || J \rangle.
\end{equation}
$\langle r^{-3} \rangle$ is the expectation value of the inverse-cube electron-nuclear distance ($\langle r^{-3} \rangle=5.0$~a.u.\ for Pr$^{3+}$ \cite{abragam77}) and:
\begin{align} \label{eq:JNJ}
\langle J || N || J \rangle &=  \frac{1}{J(J+1)} [ \left(\boldsymbol{L} \cdot \boldsymbol{J}\right) + \xi [L(L+1)(\boldsymbol{S} \cdot \boldsymbol{J}) \nonumber \\
&\quad - 3(\boldsymbol{L} \cdot \boldsymbol{J})(\boldsymbol{L} \cdot \boldsymbol{S}) ] ] ,
\end{align}
where $\xi = \frac{2\ell +1 - 4S}{S(2\ell-1)(2\ell+3)(2L-1)}$. In this work the values of the magnetic hyperfine constant given in reference \cite{abragam77} are used. These were calculated from magnetic resonance measurements on salts, but should also be applicable to crystalline hosts. 

From Eqs.~(\ref{eq:Lambda})--(\ref{eq:JNJ}) the parameters of the second-order Zeeman tensor are calculated. These are summarized in Table~\ref{tab:t_pryag4}. 

When calculating $\boldsymbol{\Lambda}$ it was necessary to make certain assumptions about the sign of the enhanced nuclear gyromagnetic ratios. Raman heterodyne spectroscopy only determines the magnitude of the enhanced nuclear gyromagnetic ratios, as inverting the principle values of the enhanced nuclear Zeeman tensor leaves the spectra unchanged. This effectively means $\Lambda_{\alpha \alpha} = k|\gamma_\alpha - \gamma_N|$ ($k = \hbar/2 g_J \mu_B A_J$), which has two solutions. For the excited state the enhanced nuclear gyromagnetic ratios are approximately the same size as $\gamma_N$; therefore, either $\Lambda_{\alpha \alpha}$ is small, or $\Lambda_{\alpha \alpha} \simeq 2 k \gamma_N$. It is assumed that the former is more likely; therefore, the sign of $\gamma_\alpha$ was chosen such that $|\gamma_\alpha - \gamma_N| < \gamma_\alpha$. The situation is different for the ground state in that $\gamma_x$ and $\gamma_z$ are significantly larger than $\gamma_N$. In this case it is difficult to make an assumption about the sign of $\gamma_\alpha$; however, either choice will result in similar values of $\Lambda_{\alpha \alpha}$, that is, $\Lambda_{\alpha \alpha} \simeq k |\gamma_\alpha|$. Here the sign of $\gamma_\alpha$ was again chosen such that $|\gamma_\alpha - \gamma_N| < \gamma_\alpha$. It is thought that the resulting landscape of \zefoz transitions will be similar for both choices of $\gamma_\alpha$, because for the magnetic fields at which \zefoz transitions exist ($10^2 - 10^4$~G) the magnitude of $ \boldsymbol{B} \cdot \boldsymbol{Z} \cdot \boldsymbol{B}$ is smaller than the magnitude of both $\boldsymbol{B} \cdot \boldsymbol{M} \cdot \boldsymbol{I}$ and $\boldsymbol{I} \cdot \boldsymbol{Q} \cdot \boldsymbol{I}$. Therefore, even with the wrong choice of $\gamma_\alpha$, it should be possible to find the \zefoz transitions experimentally by tuning the magnetic field.

If the \zefoz transitions cannot be found experimentally there are two options. The first is to calculate the \zefoz transitions for the alternative values of $\Lambda_{\alpha \alpha}$. The second, better option would be to measure the second-order Zeeman tensor, and recalculate the \zefoz transitions using this information. The second-order Zeeman tensor could be measured by burning a spectral hole, then measuring the frequency shift of the hole when a magnetic field is applied.

\begin{table}[t]
\caption{\label{tab:t_pryag4} Components of the second-order Zeeman tensor in Pr$^{3+}$:YAG.}
\begin{center}
\begin{tabular}{ccc}
\toprule
\toprule
& Ground State ($^3$H$_4$)   & Excited State ($^1$D$_2$)  \\ 
\otoprule
$g_J$  & 0.8 & 1.0  \\
 $\langle J || N || J \rangle$ & 1.316 & 1.000 \\
$A_J$ (MHz) \cite{abragam77} &  1093 & 831 \\
$Z_{x}$ (kHzG$^{-2}$) & $9.54 \times 10^{-4}$  & $1.15 \times 10^{-5}$\\
$Z_{y}$ (kHzG$^{-2}$) & $2.41 \times 10^{-5}$ & $5.03 \times 10^{-5}$ \\
$Z_{z}$ (kHzG$^{-2}$)& $2.37 \times 10^{-3}$ &  $4.37 \times 10^{-5}$\\
\bottomrule
\bottomrule
\end{tabular}
\end{center}
\end{table}

\begin{figure}[t]
 \begin{center}
  \includegraphics[width=.45\textwidth]{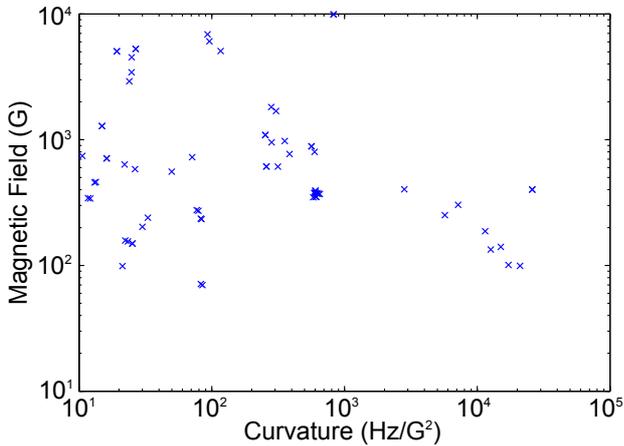}
\end{center}
\caption{\label{fig:zefozpryag} (Color online) Magnitude of the applied magnetic field versus the maximum curvature of the \zefoz transitions found for \PRYAGb. }
\end{figure}

Now that the entire spin Hamiltonian is known, for both the ground and excited states, the optical \zefoz transitions can be found. This is done by adopting a method similar to Longdell \textit{et al.} \cite{longdell06}. By searching over a three-dimensional grid of magnetic field values, transitions are found with a gradient vector equal to zero. A three-dimensional Newton-Raphson method is used to hone in on the zero points. The iteration:
\begin{equation}\label{eq:zefoza}
 \textbf{B}_\text{new} = \textbf{B} - \frac{\textbf{v}}{2 \textbf{C}},
\end{equation}
is used, where $\textbf{v}$ and $\textbf{C}$ are the Zeeman gradient and curvature tensors respectively:
\begin{equation} \label{eq:zefozb}
 v_i^{pq}(\textbf{B}) = \frac{\partial}{\partial B_i} [\omega_p(\textbf{B}) - \omega_q(\textbf{B})],
\end{equation}
\begin{equation} \label{eq:zefozc}
 C_{ij}^{pq}(\textbf{B}) = \frac{\partial^2}{\partial B_i \partial B_j} [\omega_p(\textbf{B}) - \omega_q(\textbf{B})].
\end{equation}
These are calculated using first- and second-order perturbation theory:
\begin{equation} \label{eq:zefozd}
 \frac{\partial}{\partial B_i} \omega_p(\textbf{B}) = \langle \phi_p (\textbf{B}) | \zeta_{ij}| \phi_p(\textbf{B}) \rangle,
\end{equation}
\begin{equation} \label{eq:zefoze}
 \frac{\partial^2}{\partial B_i \partial B_j} \omega_p(\textbf{B}) = \displaystyle\sum\limits_{q\neq p}\frac{\langle \phi_p (\textbf{B}) | \zeta_{ik}| \phi_q(\textbf{B}) \rangle \langle \phi_q (\textbf{B}) | \zeta_{jl}| \phi_p(\textbf{B}) \rangle}{\omega_p (\textbf{B}) - \omega_q (\textbf{B})},
\end{equation}
where $\zeta_{ij} =  Z_{ij}B_j + B_i Z_{ij} + M_{ij} I_j$.  The \zefoz transitions found for \PRYAG are shown in Fig.~\ref{fig:zefozpryag}. A total of 122 \zefoz transitions were found with curvatures ranging from $10^1$ to $10^5$~HzG$^{-2}$. The magnitude of the required magnetic fields ranges from $50$ to $10000$~G. The transition with the lowest curvature should result in the largest increase in coherence time. This is because having a lower curvature means the transition frequency is more resilient to small fluctuations in magnetic field.

In previous \zefoz experiments, operating on a turning point with curvature of 60~HzG$^{-2}$ resulted in a 1700-fold increase in the coherence time of \PRYSO \cite{longdell06}; likewise, in Pr$^{3+}$:La$_2$(WO$_4$)$_3$ a turning point with curvature of 120~HzG$^{-2}$ resulted in a 630-fold increase \cite{lovric11}. In \PRYAG only a 23-fold increase is required to achieve a population lifetime limited coherence time (from 20~$\mu$s to 460~$\mu$s), and turning points exist with curvatures as low as 10.6~HzG$^{-2}$. Therefore, by operating on a \zefoz transition we expect to see a significant increase in the optical coherence time.

\section{Er$^{3+}$:\YSO}

There are two main factors that make \ERYSO an interesting material system for studying with respect to rare-earth-ion doped quantum computing. Firstly, the $^4$I$_{15/2}$ -- $^4$I$_{13/2}$ transition occurs at 1536~nm (site 1) and 1538~nm (site 2). This is important because it falls within the telecommunications C band. For example, were a working quantum repeater to be realized from \ERYSOb, existing telecommunications infrastructure could be used, greatly reducing the cost of implementing a quantum network. Secondly, the $^4$I$_{13/2}$ excited state of \ERYSO has a long optical population lifetime ($T_{1(opt)}=11.4$~ms for site 1, $T_{1(opt)}=9.2$~ms for site 2 \cite{bottger06}), and the longest optical coherence time ($T_{2(opt)}=4.38$~ms \cite{bottger09}) of any solid-state system. The aim of this work is to find a \zefoz transition in \ERYSOb, with the intention of extending the coherence time of a hyperfine transition. This should allow long-term storage of quantum states for quantum computing applications.

There have been several investigations into the lifetime of the ground state levels in \ERYSO \cite{hastingssimon08, baldit10}. Using spectral hole burning Hastings-Simon \textit{et al.} measured the ground state lifetime between electronic Zeeman levels to be $T_{1(hyp)}=130$~ms; however, they also observed that a subset of the ions had lifetimes up to 60~s \cite{hastingssimon08}. Baldit \textit{et al.} measured the hyperfine lifetime of isotopically pure $^{167}$Er to be $T_{1(hyp)}=97$~ms \cite{baldit10}. Based on the inequality $T_2 \leq 2 T_1$ it should be possible to attain hyperfine coherence times up to $T_{2(hyp)}=194$~ms.

Erbium has six stable isotopes of which only $^{167}$Er has a non-zero value of nuclear spin. This isotope is of particular interest as the non-zero nuclear spin results in energy levels that exhibit hyperfine structure. $^{167}$Er has a natural abundance of 22.95\% \cite{laeter03} and a nuclear spin of $I=\frac{7}{2}$. For the $J=\frac{15}{2}$ ground and $J=\frac{13}{2}$ excited states the crystal field splits into eight and seven Kramers' doublets respectively. As stated earlier the system is approximated as having spin $S = \frac{1}{2}$; thus each crystal-field level will split into a total of sixteen hyperfine levels (two electronic spin states (m$_\text{S}=\frac{1}{2}$, $\frac{-1}{2}$) $\times$ eight nuclear spin states (m$_\text{I}=\frac{-7}{2}$ -- $\frac{7}{2}$)). 
                                                                                                                                                                                                                                                                                                                                                                                                                                                                                                                                                                                                               
Using electron paramagnetic resonance spectroscopy, $\textbf{g}$, $\textbf{A}$, and $\textbf{Q}$ have been measured by Guillot-No\"el \textit{et al.} for the $^4$I$_{15/2}$ ground state of sites 1 and 2 \cite{guillot-noel06}. Sun \textit{et al.} used optical Zeeman spectroscopy to measure $\textbf{g}$ for the ground and excited states of both sites \cite{sun08}. These two separate measurements of $\textbf{g}$ for the ground state agree extremely well, except for along the principle $x$-axis of site 1. This discrepancy is because the magnitude of $\textbf{g}$ in the $x$ direction is significantly smaller than in the other directions; therefore, the measurements are less sensitive to variation along this axis. For the calculations performed in this work, the parameters measured by Guillot-No\"el \textit{et al.} \cite{guillot-noel06} are used as they constitute a complete measurement of the ground-state spin Hamiltonian. For site 1:
\[ \textbf{g}_1 = \begin{bmatrix}
 2.92 & -3.08 & -3.68 \\
 -3.08 & 8.19 & 5.96 \\
 -3.68 & 5.96 & 5.52 \end{bmatrix},\] 
\[ \textbf{A}_1 = \begin{bmatrix}
 69.35 & -580.73 & -248.83 \\
 -580.73 & 696.30 & 682.49 \\
 -248.83 & 682.49 & 495.54 \end{bmatrix} \text{MHz},\] 
\[ \textbf{Q}_1 = \begin{bmatrix}
 21.40 & -8.18 & -15.27 \\
 -8.18 & 3.79 & 0.60 \\
 -15.27 & 0.60 & -25.20 \end{bmatrix} \text{MHz},\]
and for site 2:
\[ \textbf{g}_2 = \begin{bmatrix}
   14.75 & -2.02  & 2.62  \\
  -2.02 & 1.89 & -0.93  \\
  2.62 & -0.93  & 0.05  \end{bmatrix},\] 
\[ \textbf{A}_2 = \begin{bmatrix}
  -1521.40 & 178.11  & -141.76  \\
  178.11 & 172.09 & 212.54 \\
 -141.76 & 212.54 & 199.01 \end{bmatrix} \text{MHz},\] 
\[ \textbf{Q}_2 = \begin{bmatrix}
 -3.50 & -19.84 & 24.22 \\
 -19.84 & 50.40 & 6.73 \\
 24.22 & 6.73 & -46.90 \end{bmatrix} \text{MHz}.\]
The coordinates are defined in terms of the axes of the \YSO crystal, that is, $x=D_1$-axis, $y=D_2$-axis, $z=b$-axis.

\begin{figure}[t]
  \begin{center}
  \includegraphics[width=.45\textwidth]{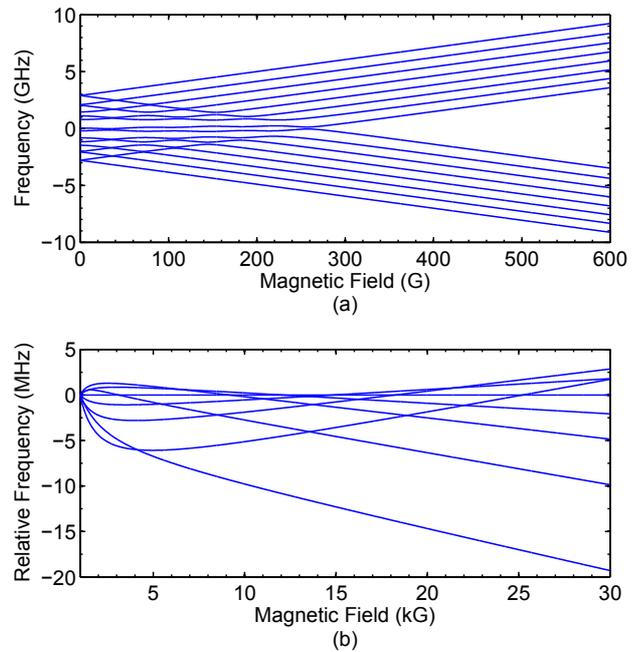}
\end{center}
\caption{\label{fig:fieldvfreq} (Color online) Energy of the site 2 hyperfine levels as a magnetic field is applied along the $D_1$-axis of the crystal for; (a) low magnetic fields, and (b) high magnetic fields. To gain an understanding of the relative frequency between levels at high fields, in (b) the frequency of the lowest eight energy levels relative to the 4th lowest energy level are plotted. For clarity an extra offset has been added to each level such that the relative frequency is zero at $B=1$~kG. (Note that both the $x$ and $y$ axes for the graphs are different.)
}
\end{figure}

Knowledge of the entire spin Hamiltonian allows the frequency of the hyperfine levels to be calculated for any applied magnetic field. As an example, Fig.~\ref{fig:fieldvfreq}(a) shows how the ground-state hyperfine splittings change as a magnetic field is applied along the $D_1$-axis of the crystal. Notice that the energy levels exhibit a large number of anticrossings; therefore, \zefoz transitions should exist. At large applied fields the electronic Zeeman interaction dominates and the system displays two distinct levels corresponding to the two electronic states, each of which is split into eight hyperfine levels. 

To gain an understanding of why \zefoz transitions exist even for large magnetic fields the relative frequency of the $m_{S}=-1/2$ hyperfine levels is plotted in Fig.~\ref{fig:fieldvfreq}(b). The relative fluctuations in transition frequency are small (on the order of a few MHz) which will result in \zefoz transitions with a low curvature.

\subsection{Finding the \zefoz transitions}

\begin{figure}[t]
  \begin{center}
  \includegraphics[width=.45\textwidth]{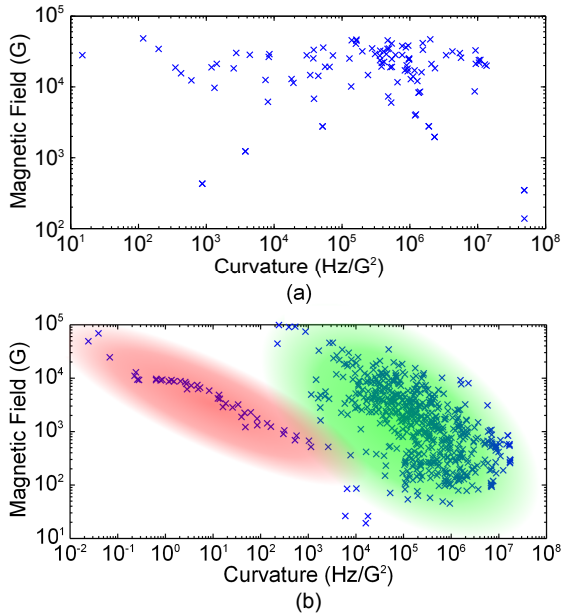}
\end{center}
\caption{\label{fig:bvcurv} (Color online)
Magnitude of the applied magnetic field versus the maximum curvature of the \zefoz transitions found for; (a) site 1, and (b) site 2 of \ERYSOb. Note the two shaded regions in (b) which correspond to the two different regimes as described in the text.
 }
\end{figure}

Similar to the earlier work on Pr$^{3+}$:YAG, a three-dimensional Newton-Raphson method is used to find the hyperfine \zefoz transitions in \ERYSOb. The gradient and curvature are as defined in Eqs.~(\ref{eq:zefozb})--(\ref{eq:zefoze}) except:
\begin{equation}
\zeta_{ij} =  \mu_B g_{ij}S_j - \mu_n g_n I_i.
\end{equation}
Figure~\ref{fig:bvcurv} is a plot of the absolute value of the applied magnetic field versus the maximum value of the curvature for the identified \zefoz transitions. For site 1 a total of 102 \zefoz transitions were found, with curvatures ranging from $10^1$ to $10^8$~HzG$^{-2}$ and requiring magnetic fields ranging from $0.1$ to $100$~kG. 635 \zefoz transitions were found for site 2, with curvatures ranging from $10^{-2}$ to $10^8$~HzG$^{-2}$. The required magnetic fields range from $0.01$ to $100$~kG.

The maximum curvature of the best site 2 \zefoz transitions, were three orders of magnitude smaller than for those of site 1. Therefore, when experimentally trying to measure extended coherence times it is sensible to focus on site 2, as the transition with the lowest curvature should give the largest increase in coherence time. 

\subsection{Properties of the \zefoz transitions}

In Fig.~\ref{fig:bvcurv}(b) there are two different shaded regions. The shaded green (right) region corresponds to \zefoz transitions in the `strongly mixed' regime, whilst the shaded red (left) region corresponds to the `weakly mixed' regime. The strongly mixed regime is where the electronic Zeeman splitting (the term  $\mu_B \textbf{B} \cdot \textbf{g}\cdot \textbf{S}$ in Eq.~\ref{eq:erham}) is either small or comparable to the hyperfine splitting ($\textbf{I}\cdot \textbf{A} \cdot \textbf{S}$). The weakly mixed regime is where the electronic Zeeman term dominates and the electron spin is close to a good quantum number. Figure~4 shows the energy levels for site~2 as the field is increased along the crystal $D_1$-axis. For low fields ($<300\,$G for this field direction) there are many avoided crossings due to the interplay between the hyperfine splitting and the electronic Zeeman term. It is in this area where the strongly mixed \zefoz transitions occur. At high fields where the electronic Zeeman term is larger than the hyperfine splitting the electronic spin becomes close to a good quantum number and Fig.~4(a) shows the energy levels split into two branches each with a different electron spin. The energy levels in each branch look like they move together as the applied field is changed. Figure~4(b) shows that this is only roughly the case and what look like straight parallel lines do have some curvature and structure. It is in this area where the weakly mixed \zefoz transitions occur.

The transitions in the weakly mixed regime can be expected to provide the longest coherence times as in general they are of lower curvature than those in the strongly mixed regime. The transition strength of the strongly mixed transitions will be much larger as the \zefoz transitions in the weakly mixed regime are essentially nuclear spin transitions. Even though the strongly mixed regime is characterised by small Zeeman splittings compared to hyperfine splittings, a reasonable number of these transitions occur at relatively high fields (>1~kG). This is because $\textbf{g}$ is highly anisotropic, meaning that there are certain field directions for which $\mu_B \textbf{B} \cdot \textbf{g}\cdot \textbf{S}$ is of comparable magnitude to $\textbf{I}\cdot \textbf{A} \cdot \textbf{S}$ resulting in energy levels which are still highly mixed.

To obtain an estimate of the coherence times ($T_2$) that should be attainable in \ERYSO the dephasing can be simply modelled as \cite{longdell06}:
\begin{equation}
\frac{1}{T_2} = S_2 (\Delta B)^2,
\end{equation}
where $S_2$ is the curvature of the \zefoz transition and $(\Delta B)^2$ is the variance of the magnetic field fluctuations. From reference \cite{longdell06} the magnetic field fluctuations experienced by praseodymium ions in \YSO are estimated as $(\Delta B)^2 =  0.0196$~G$^2$. For erbium dopants this could be considerably smaller due to the larger magnetic moment of erbium inducing a much larger frozen core and so lifetime limited coherence times of the order of $200$~ms are a possibility. In order to get the very long coherence times promised by the much flatter weakly mixed ZEFOZ transitions one needs to get around the problem of much faster spin lattice relaxation rates at high fields \cite{hastingssimon08}. One way to solve this problem would be to operate at temperatures smaller than the splitting between the two branches and use a ZEFOZ transition in the lower branch. In this way the electron spin would always be in its lowest energy state and its relaxation rate would be immaterial. 

\subsection{Transition Strengths and Hyperfine State Mixing}

More than 700 \zefoz transitions have been found in sites 1 and 2 of \ERYSOb; however, an important consideration is how easy it is to excite the transition either directly using an RF field or, of particular importance for quantum memories, via Raman transitions using the bottom level of the $^4I_{13/2}$ excited multiplet. Precise values for the Raman transition strengths are not currently available because the excited state spin Hamiltonian is not known. However, arguments about what transitions are likely to be strong and what are likely to be weak can still be made. For \zefoz transitions in the strongly mixed regime the mixing greatly relaxes the spin selection rules, allowing $\Lambda$-systems to be formed. This is especially the case around avoided crossings, which is where many of the \zefoz transitions occur. An illustration of this is provided by Baldit \textit{et al.} where many $\Lambda$-systems were found at zero magnetic field \cite{baldit10}.

For the weakly mixed regime the electronic Zeeman term dominates and the rest can be treated as a perturbation. This means that $m_{S}$ is a good quantum number and the electron spin states are simply angular momentum eigenstates with the quantization direction given by $\mathbf{B}\cdot\mathbf{g}$. For each value of $m_{S}$ there are eight hyperfine levels as the hyperfine contribution ($\mathbf{A}$) is large  compared to the nuclear quadrupole ($\mathbf{Q}$), at least up to very high fields where the nuclear Zeeman term becomes important. By projecting the hyperfine term onto one of the $m_S=\pm1/2$ branches it becomes of the form $\pm\mathbf{I}\cdot \mathbf{A} \cdot (1/2 \mathbf{\hat{n}})$, where the unit vector $\mathbf{\hat{n}}$ is the quantization direction for the electron spin. So like the electron spin, the dominant effect in the nuclear spin splitting for each of the two  branches has the same form as a Zeeman splitting. This leads to nuclear spin states which are also angular momentum eigenstates, but with a quantization axis that is in general different to that for the electron spin. So in the weakly mixed regime the transitions that leave the electron spin unchanged will be allowed and can be driven with a RF magnetic field so long as the usual $\Delta m_I=\pm 1$ condition is satisfied. 

The nuclear spin contribution to the transition strength of an optical transition is simply the overlap between the ground state and excited state nuclear spin states. If the quantization axes are the same for the nuclear spins in the ground and excited states then Raman transitions will not be possible because the nuclear spin states will only overlap if the two $m_I$ values are the same. However, because the electronic Zeeman ($\mathbf{g}$) and hyperfine ($\mathbf{A}$) tensors are anisotropic, and are different for the ground and excited states, the quantization axes will be different. The extent to which a Raman transition can be found will depend on the angle between the quantization axes. For small angle differences the change of basis matrix between the ground state and excited state energy levels will be clustered around the diagonal, making Raman transitions possible only for small values of $\Delta m_I$. 

Thus as an alternative to calculating the optical transition strengths, which would require knowledge of the spin Hamiltonian for the excited state, the values  for $|\Delta m_I|$ are given. Whilst this does not explicitly determine whether it will be possible to create a $\Lambda$-system from the transition, it does give some indication. If $\Delta m_I$ is small the chances of finding a strong $\Lambda$-system are much higher.

The strengths of the \zefoz transitions, when driven with an RF field, can be calculated directly from the spin Hamiltonian using the probability amplitude method \cite{scully01}. It is assumed that the transition is being driven by an AC magnetic field (B$_{\text{AC}}$) oscillating at $\omega$ (where $\omega$ is the transition frequency). This situation can be described by the interaction picture Hamiltonian:
\begin{equation}
 H = H_0 + H_\text{I},
\end{equation}
where $H_0$ is the spin Hamiltonian of Eq. (\ref{eq:erham}) and can be written as:
\begin{align}
H_0 & =  (\ket{a}\bra{a} + \ket{b}\bra{b}) H_0 (\ket{a}\bra{a} + \ket{b}\bra{b}) \nonumber \\ 
  & =  \hbar \omega_a \ket{a} \bra{a} + \hbar \omega_b \ket{b} \bra{b},
\end{align}
where $\omega_a$, $\omega_b$ are the energies of the levels involved in the transition. $H_\text{I}$ represents the interaction of the atoms with the field:
\begin{align}
 H_\text{I} & =  \text{B}_{\text{AC}} \text{cos}(\omega t) [\mu_B \textbf{g} \cdot \textbf{S} + \mu_n g_n \textbf{I}] \nonumber \\ 
    & =  \text{B}_{\text{AC}} \text{cos}(\omega t) (\ket{a}\bra{a} + \ket{b}\bra{b})[ \mu_B \textbf{g} \cdot \textbf{S} \nonumber \\  
    & \quad + \mu_n g_n \textbf{I}] (\ket{a}\bra{a} + \ket{b}\bra{b}) \nonumber \\ 
    & =  \text{B}_{\text{AC}} \text{cos}(\omega t) [\rho_{aa} \ket{a}\bra{a} + \rho_{ab} \ket{a}\bra{b} \nonumber \\
    & \quad + \rho_{ba}\ket{b}\bra{a} + \rho_{bb}\ket{b}\bra{b}].
\end{align}
 $\rho_{ab} = \bra{a}\mu_B \textbf{g} \cdot \textbf{S} + \mu_n g_n \textbf{I}\ket{b}$ is the magnetic dipole moment of the transition from $\ket{a}$ to $\ket{b}$. 

As a measure of the transition strength the magnetic dipole moment of the transitions has been calculated. The magnitude of this is listed in Table~\ref{tab:t_erysotransitions3} for specific \zefoz transitions. The \zefoz transitions with the lowest curvatures tend to have weaker transition strengths than those with higher curvatures. However, by making a compromise between curvature and transition strength, several transitions have been found that have the potential to enable measurement of extended coherence times. Table~\ref{tab:t_erysotransitions3} lists the properties of \zefoz transitions with curvatures $\sim 10^3$~HzG$^{-2}$ and transition strengths $> 10^{5}$~HzG$^{-1}$.

\begin{table*}[t]
\caption{\label{tab:t_erysotransitions3} \zefoz transitions in \ERYSO where a compromise has been made between transition strength and maximum curvature. For all of these transitions $|\Delta m_I|=1$. The levels are labelled 1 - 16, from lowest to highest frequency as depicted in Fig.~\ref{fig:fieldvfreq}(a).}
\begin{center}
\newcommand{\minitab}[2][l]{\begin{tabular}{#1}#2\end{tabular}}
\begin{tabular}{ccccccccccccc}
\toprule
\toprule
\multirow{2}{*}{Site}&\multirow{2}{*}{Level 1} &  \multirow{2}{*}{Level 2}   & Maximum &   \multirow{2}{*}{B$_{D_1}$ (kG)} &   \multirow{2}{*}{B$_{D_2}$ (kG)} &   \multirow{2}{*}{B$_{b}$ (kG)} &   \multirow{2}{*}{|B| (kG)} & Transition & Transition\\ 
&& &  Curvature (HzG$^{-2}$) & & & & & Strength (kHzG$^{-1}$)& Frequency (MHz) \\
\otoprule
1&7 & 8 & 1193 & 11.43 & -6.136 & 13.73 & 18.89 & 279.1 & 905.7 \\
2&10 & 11 & 225.7 & -7.992 & 3.529 & 43.36 & 44.23 & 361.9 & 408.0 \\
2&7 & 8 & 392.7 & -13.30 & 9.169 & 88.78 & 90.23 & 243.0 & 239.8 \\
2&11 & 12 & 1153 & -1.029 & -4.825 & 0.2901 & 4.942 & 887.0 & 324.0 \\
2&12 & 13 & 1458 & -2.264 & -6.259 & 6.567 & 9.350 & 427.0 & 105.9 \\
2&5 & 6 & 1615 & 0.5843 & 4.024 & 1.173 & 4.232 & 1292 & 322.2 \\
2&6 & 7 & 1790 & -0.5366 & 0.2861 & 2.748 & 2.815 & 2796 & 183.8 \\
2&6 & 7 & 1805 & -0.2467 & -0.7551 & 1.411 & 1.619 & 4956 & 465.2 \\
2&12 & 13 & 1812 & -0.7285 & 0.09411 & 4.776 & 4.832 & 4276 & 417.9 \\
\bottomrule
\bottomrule
\end{tabular}
\end{center}
\end{table*}

\section{Conclusion}

In this work the \zefoz transitions for optical transitions in \PRYAG and ground-state hyperfine transitions in \ERYSO were calculated. This shows that the ZEFOZ technique which has previously only been used on the hyperfine transitions of non-Kramers' ions is also applicable to optical transitions and also to Kramers' ions.

In \PRYAG the terms of the excited-state enhanced nuclear Zeeman tensor were measured using Raman-heterodyne spectroscopy. The second-order Zeeman tensor was then calculated for the ground and excited states. Based on these parameters and those measured by Wang, the location of the \zefoz transitions were determined. 

In \ERYSO the location of \zefoz transitions for both sites were calculated based on the spin Hamiltonian parameters measured by Guillot-No\"el \textit{et al.} Transitions were found for site 2 that have very low curvatures; therefore, it should be possible to significantly lengthen the hyperfine coherence time. The strength of the hyperfine transitions was calculated, and it was determined that \zefoz transitions exist that simultaneously have low curvatures and high transition strengths.

\section{Acknowledgments}

The authors are grateful to A. Louchet-Chauvet, J.-L. Le Gou\"et, P. Goldner, C. W. Thiel and R. L. Cone for useful discussions on the properties of \ERYSO

DLM and JJL were supported by the New Zealand Foundation for Research Science and Technology under Contract No.\ NERF-UOOX0703. JGB and MJS were supported by the Australian Research Council Centre of Excellence for Quantum Computation and Communication Technology (Project No. CE11E0096).


\end{document}